\documentclass{revtex4}
\usepackage{amssymb}
\usepackage{amsfonts}
\usepackage{amsmath}

\begin{document}

\title{Comment on 'Two-dimensional position-dependent massive particles in
the presence of magnetic fields"}
\author{Omar Mustafa}
\email{omar.mustafa@emu.edu.tr}
\affiliation{Department of Physics, Eastern Mediterranean University, G. Magusa, north
Cyprus, Mersin 10 - Turkey,\\
Tel.: +90 392 6301378; fax: +90 3692 365 1604.}

\begin{abstract}
\textbf{Abstract:} Using the well known position-dependent mass (PDM) von
Roos Hamiltonian, Dutra and Oliveira (2009 J. Phys. \textbf{A}: Math. Theor. 
\textbf{42} 025304) have studied the problem of two-dimensional PDM
particles in the presence of magnetic fields. They have reported exact
solutions for the wavefunctions and energies. In the first part of their
study "PDM-Shr\"{o}dinger equation in two-dimensional Cartesian
coordinates", they have used the so called Zhu and Kroemer's ordering $%
\alpha =-1/2=\gamma $ and $\beta =0$ \cite{5}. While their treatment for
this part is correct beyond doubt, their treatment of second part "PDM in a
magnetic field" is improper . We address these improper treatments and
report the correct presentation for minimal coupling under PDM settings.


\end{abstract}

\maketitle

Over the last few decades, both classical and quantum mechanical particles
endowed with position-dependent mass (PDM) have attracted much research
attention (c.f., e.g., \cite{1,2,3,4,5,6,7,8} and references cited therein).
The study of a classical PDM-nonlinear oscillator by Mathews and Lakshmanan 
\cite{8} has sparked and inspired a large number of research in different
fields of study. It was only recently, to the best of our knowledge, that
the effect of a uniform magnetic field on a PDM quantum particle in two
dimensions is studied by Dutra and Oliveira \cite{1}. They have reported
exact wavefunctions and energies for such a problem. Whilst the first part
of their "PDM-Shr\"{o}dinger equation in two-dimensional Cartesian
coordinates" follows Zhu and Kroemer's ordering $\alpha =-1/2=\gamma $ and $%
\beta =0$ \cite{5} (their Eq. (24)) and is correct beyond doubt, their
second part "PDM in a magnetic field" suffers improper and conflicting
treatments that inspired the current obligatory comments.

In their attempt to analyze and discuss the quantum mechanical effect of a
uniform magnetic field on a charged particle endowed with position-dependent
mass in two dimensions, Dutra and Oliveira \cite{1} have started (in their
section 3) with the classical Hamiltonian 
\begin{equation}
H=\frac{1}{2M\left( x,y\right) }\left( \overrightarrow{p}-e\,\overrightarrow{%
A}\right) ^{2}+V\left( x,y\right) =\frac{1}{2M}\overrightarrow{p}^{2}-\frac{e%
}{2M}\left( \overrightarrow{A}\cdot \overrightarrow{p}+\overrightarrow{p}%
\cdot \overrightarrow{A}\right) +\frac{e^{2}}{2M}\overrightarrow{A}%
^{2}+V\left( x,y\right) .
\end{equation}%
Where $e$ is the electric charge, $M\equiv M\left( x,y\right) $ is the PDM, $%
\overrightarrow{A}\equiv \overrightarrow{A}\left( x,y\right) $ is the vector
potential, and $V\left( x,y\right) $ is a scalar potential (their equation
(37)). Next, with the substitution ((38) in \cite{1})%
\begin{equation}
\overrightarrow{\tilde{A}}=\frac{\overrightarrow{A}\left( x,y\right) }{%
M\left( x,y\right) }=\frac{\overrightarrow{A}}{M},
\end{equation}%
they have obtained the classical Hamiltonian%
\begin{equation}
H=\frac{1}{2M}\overrightarrow{p}^{\,2}-\frac{e}{2}\left( \overrightarrow{%
\tilde{A}}\cdot \overrightarrow{p}+\overrightarrow{p}\cdot \overrightarrow{%
\tilde{A}}\right) +\frac{e^{2}M}{2}\overrightarrow{\tilde{A}}^{2}+V\left(
x,y\right) .
\end{equation}%
At this point, they have suggested that the ordering of the kinetic energy
operator is discussed in their section 2 and hence the terms without
magnetic interaction (i.e.,\ $\vec{A}\left( x,y\right) =0$) in (3) are given
by their Eq.(9) as%
\begin{equation}
\frac{1}{2M}\overrightarrow{p}^{\,2}+V\left( x,y\right) =-\frac{\hbar ^{2}}{%
2M}\overrightarrow{\nabla }^{\,2}+\frac{\hbar ^{2}}{2M^{2}}\left( 
\overrightarrow{\nabla }M\cdot \overrightarrow{\nabla }\right) +U\left(
\alpha ,\gamma ,x,y\right) +V\left( x,y\right) ,
\end{equation}%
\newline
where $U\left( \alpha ,\gamma ,x,y\right) $ is given by their Eq.(8). But
then, to deal with the operator linear in $\overrightarrow{p}$ (i.e., $%
\overrightarrow{\tilde{A}}\cdot \overrightarrow{p}+\overrightarrow{p}\cdot 
\overrightarrow{\tilde{A}}$, the 2nd term in (3) or their 2nd term of (39))
they have used $\overrightarrow{p}$ $=-i\hbar \overrightarrow{\nabla }$ to
obtain their Eq.(50). Next, they used $\psi \left( x,y\right) =M^{1/2}\chi
\left( x,y\right) $ in (50) to get their Eq.(51) and proceeded with their
solution considering the well known Zhu and Kroemer's ordering $\alpha
=-1/2=\gamma $ and $\beta =0$ \cite{5} (to get rid of the differential forms
of the PDM terms).

In the light of our experience and practical contact with this paper \cite{1}
, we feel obligated to pin point our observations that are in order:

1- In handling (3), they have used two conflicting/inconsistent definitions
for the momentum operator. Having used%
\begin{equation}
\frac{1}{2M}\overrightarrow{p}^{\,2}=-\frac{\hbar ^{2}}{2M}\overrightarrow{%
\nabla }^{\,2}+\frac{\hbar ^{2}}{2M^{2}}\left( \overrightarrow{\nabla }%
M\cdot \overrightarrow{\nabla }\right) +U\left( \alpha ,\gamma ,x,y\right) 
\end{equation}%
in (4) (their (39) to obtain (50)) necessarily suggests that $%
\overrightarrow{p}$ $\neq -i\hbar \overrightarrow{\nabla \text{ }}$ (as
documented in their Eq.s (43) to (46)). One should notice that $%
\overrightarrow{p}$ in (1) is the PDM canonical momentum and should
correspond to PDM-momentum operator in (4). In fact, by a factorization
recipe \cite{3}, equation (5) immediately implies that the PDM-momentum
operator should look like that in their equation (42) as%
\begin{equation}
\widehat{O}=f\left( x\right) \widehat{p}-\frac{i\hbar }{2}\left( \frac{%
df\left( x\right) }{dx}\right) \Longrightarrow \widehat{O}^{2}=\frac{1}{2M}%
\widehat{p}^{\,2}\Longrightarrow f\left( x\right) =\frac{1}{\sqrt{2M}}.
\end{equation}%
Which will, consequently, strictly determine the ordering parameters as $%
\alpha =\gamma =-1/4$ and $\beta =-1/2$ (known as MM-ordering \cite{3}).
Moreover, it suggests that 
\begin{equation}
\widehat{O}=\frac{\widehat{p}}{\sqrt{2M}}=\frac{-i\hbar }{\sqrt{2M}}\left[ 
\frac{d}{dx}-\frac{1}{4}\frac{\left( dM/dx\right) }{M}\right]
\Longrightarrow \widehat{p}_{x}=-i\hbar \left[ \frac{\partial }{\partial x}-%
\frac{1}{4}\frac{\left( \partial M/\partial x\right) }{M}\right]
\Longrightarrow \overrightarrow{p}_{op}=-i\hbar \left[ \overrightarrow{%
\nabla }-\frac{\left( \overrightarrow{\nabla }M\right) }{4M}\right] .
\end{equation}%
Which, in turn, collapses into $\overrightarrow{p}_{op}=-i\hbar 
\overrightarrow{\nabla }$ for constant mass settings as it should.

2- The von Roos operator they have used (their Eq.(1)) as%
\begin{equation}
\hat{H}=\frac{1}{4}\left[ M^{\alpha }\widehat{p}\,M^{\beta }\widehat{p}%
M^{\gamma }+M^{\gamma }\widehat{p}M^{\beta }\widehat{p}M^{\alpha }\right]
+V\left( x,y\right) ,
\end{equation}%
caused all the confusion, in our opinion. In the von Roos \cite{2} proposal,
this Hamiltonian operator is given in a differential\ form as

\begin{equation}
\hat{H}=-\frac{1}{4}\left[ M^{\alpha }\partial _{x_{j}}M^{\beta }\partial
_{x_{j}}M^{\gamma }+M^{\gamma }\partial _{x_{j}}M^{\beta }\partial
_{x_{j}}M^{\alpha }\right] +V\left( x,y\right) ,\text{ \ }j=1,2
\end{equation}%
where the summation runs over the repeated index, $\partial
_{x_{j}}=\partial /\partial x_{j}$\ and the ordering parameters satisfy the
von Roos constraint $\alpha +\beta +\gamma =-1$ (to recover the constant
mass settings when $M\left( x,y\right) =m_{\circ }$). However, one should be
aware that the canonical PDM-momentum $\overrightarrow{p}$ in (1) would lead
to a PDM-momentum operator, as discussed in point 1 above.. To make this
point more clear, perhaps it is wise to go back to the very fundamentals of 
\textit{Quantum mechanics} by S. Gasiorowicz \cite{9} and recollect that,
the canonical momentum for a constant mass is $p_{_{x}}=m_{\circ }\dot{x}=$ $%
m_{\circ }dx/dt$ and the corresponding quantum mechanical momentum operator
is determined through 
\begin{equation}
\left\langle p_{_{x}}\right\rangle =m_{\circ }\frac{d}{dt}\left\langle
x\right\rangle =\int\limits_{-\infty }^{\infty }dx\Psi ^{\ast }\left(
x,t\right) \left( -i\hbar \frac{\partial }{\partial x}\right) \Psi \left(
x,t\right) \Longrightarrow \widehat{p}_{x}=-i\hbar \partial /\partial x.
\end{equation}%
Whereas, the canonical PDM-momentum for a classical PDM-Lagrangian%
\begin{equation*}
L=\frac{1}{2}M\left( x,y\right) \,\dot{x}_{j}^{2}-V\left( x,y\right) 
\end{equation*}%
is given by 
\begin{equation}
P_{j}\left( x,y\right) =\partial L/\partial \dot{x}_{j}=M\left( x,y\right) 
\dot{x}_{j},
\end{equation}%
and hence the PDM-momentum operator would be found through the same recipe as%
\begin{equation}
\left\langle P_{j}\left( x,y\right) \right\rangle =\left\langle M\left(
x,y\right) \dot{x}_{j}\right\rangle \neq M\left( x,y\right) \frac{d}{dt}%
\left\langle x_{j}\right\rangle \Longrightarrow \widehat{P}_{j}\left(
x,y\right) \neq -i\hbar \partial /\partial x_{j}.
\end{equation}

3- As long as classical mechanics is concerned, the presentation of the
kinetic energy term in (1) is correct. However, when quantum mechanics is in
point, this presentation is improper. To find the proper presentation one
would recollect that the PDM Lagrangian with magnetic interaction is given by%
\begin{equation}
L\left( x_{j},\dot{x}_{j},t\right) =\frac{1}{2}M\left( x,y\right) \,\dot{x}%
_{j}^{2}+e\,\dot{x}_{j}A_{j}\left( x,y\right) -V\left( x,y\right) ,
\end{equation}%
to imply the PDM canonical momentum%
\begin{equation}
P_{j}\left( x,y\right) =\partial L/\partial \dot{x}_{j}=M\left( x,y\right) 
\dot{x}_{j}+eA_{j}\left( x,y\right) =\sqrt{M\left( x,y\right) }\Pi
_{j}\left( x,y\right) +eA_{j}\left( x,y\right) ,
\end{equation}%
where $\Pi _{j}\left( x,y\right) =\sqrt{M\left( x,y\right) }\dot{x}_{j}$ is
the $j$the component of a pseudo-mechanical-momentum (which is shown to be a
conserved quantity for a quasi-free PDM case \cite{4}, i.e., for $V\left(
x,y\right) =0=A_{j}\left( x,y\right) $). This would consequently imply that%
\begin{equation}
\Pi _{j}\left( x,y\right) =\frac{P_{j}\left( x,y\right) -eA_{j}\left(
x,y\right) }{\sqrt{M\left( x,y\right) }}.
\end{equation}%
Therefore, the corresponding classical PDM-Hamiltonian reads%
\begin{equation}
H\left( x_{j},\dot{x}_{j},t\right) =\frac{1}{2}M\left( x,y\right) \,\dot{x}%
_{j}^{2}+V\left( x,y\right) =\frac{1}{2}\Pi _{j}\left( x,y\right)
^{2}+V\left( x,y\right) =\frac{1}{2}\left( \frac{P_{j}\left( x,y\right)
-eA_{j}\left( x,y\right) }{\sqrt{M\left( x,y\right) }}\right) ^{2}+V\left(
x,y\right) .
\end{equation}%
Of course, it looks the same as that in (1) for a classical particle but
this presentation is the one to be used for a PDM quantum particle in a
magnetic interaction, with $\widehat{P}_{j}\left( x,y\right) $ as the $j$the
component of the PDM-momentum operator (7) and takes the differential form 
\cite{6} 
\begin{equation}
\widehat{P}_{j}\left( x,y\right) =\sqrt{M\left( x,y\right) }\widehat{\Pi }%
_{j}\left( x,y\right) =-i\hbar \left[ \partial _{x_{j}}-\frac{1}{4}\left( 
\frac{\partial _{x_{j}}M\left( x,y\right) }{M\left( x,y\right) }\right) %
\right] .
\end{equation}%
Under such PDM-settings, the minimal coupling for the PDM-Schr\"{o}dinger
Hamiltonian should look like 
\begin{equation}
\widehat{P}_{j}\left( x,y\right) \longrightarrow \widehat{P}_{j}\left(
x,y\right) -eA_{j}\left( x,y\right) \Longrightarrow \left( \frac{\widehat{P}%
_{j}\left( x,y\right) }{\sqrt{M\left( x,y\right) }}\right)
^{2}\longrightarrow \left( \frac{\widehat{P}_{j}\left( x,y\right)
-eA_{j}\left( x,y\right) }{\sqrt{M\left( x,y\right) }}\right) ^{2}
\end{equation}%
Hence, their equation (37) should correspond to a PDM-Schr\"{o}dinger
Hamiltonian%
\begin{equation}
\widehat{H}=\frac{1}{2}\left( \frac{\widehat{P}_{j}\left( x,y\right)
-eA_{j}\left( x,y\right) }{\sqrt{M\left( x,y\right) }}\right) ^{2}+V\left(
x,y\right) .
\end{equation}%
Quantum mechanically speaking, this Hamiltonian is not the same as the one
they have used in (1).

4- In connection with the above mentioned points, their Eq.(39) should look
exactly like%
\begin{eqnarray}
\widehat{H} &=&\frac{1}{2}\left\{ \left( \frac{\widehat{P}_{j}\left(
x,y\right) }{\sqrt{M\left( x,y\right) }}\right) ^{2}+\left( \frac{%
eA_{j}\left( x,y\right) }{\sqrt{M\left( x,y\right) }}\right) ^{2}-\left( 
\frac{eA_{j}\left( x,y\right) }{M\left( x,y\right) }\right) \widehat{P}%
_{j}\left( x,y\right) \right.  \notag \\
&&\left. -\left( \frac{\widehat{P}_{j}\left( x,y\right) }{\sqrt{M\left(
x,y\right) }}\right) \left( \frac{eA_{j}\left( x,y\right) }{\sqrt{M\left(
x,y\right) }}\right) \right\} +V\left( x,y\right) .
\end{eqnarray}%
for the quantum mechanical treatment. This is not the same as the
Hamiltonian operator they have used in (3) (their (39)).

5- Yet, in-between their (41) and (42) they have used a constraint $\alpha
+\beta =1$ to obtain (42). This would, in turn, imply that $\gamma =-2$
(using the von Roos constraint $\alpha +\beta +\gamma =-1)$. However, they
have used $\alpha =\gamma =-1/2$ and $\beta =0$ (i.e. Zhu and Kroemer
ordering \cite{5}) to obtain (54). Using two parametric orderings at the
same time is confusing and improper. Two different parametric orderings
necessarily mean two different kinetic energy operator presentations, as
obvious from (9) above.

Finally, we believe that it is necessary and vital to put PDM quantum
particles in magnetic field on the most proper track. This will open a new
window for PDM research community to make the relevant theoretical progress
not only for PDM in magnetic fields but also for PDM in electromagnetic and
laser fields.


\begin{thebibliography}{9}
\bibitem{1} A. de Souza Dutra, J A de Oliveira, J. Phys. \textbf{A}: Math.
Theor. \textbf{42,} (2009) 025304.

\bibitem{2} O. Von Roos, Phys. Rev. \textbf{B 27} (1983) 7547.

\bibitem{3} O. Mustafa, S. H. Mazharimousavi, Int. J. Theor. Phys. \textbf{46%
} (2007) 1786.

\bibitem{4} O. Mustafa, J. Phys. \textbf{A}: Math. Theor. \textbf{46,}
(2013) 368001.

\bibitem{5} Q. G. Zhu, H Kroemer, Phys. Rev. \textbf{B 27} (1983) 3519.

\bibitem{6} O. Mustafa, Z. Algadhi, arXiv:1806.02983:\textbf{\ }%
Position-dependent mass momentum operator and minimal coupling: point
canonical transformation and isospectrality.

\bibitem{7} C. Quesne, J. Math. Phys. \textbf{59} (2018) 042104.

\bibitem{8} P. M. Mathews, M. Lakshmanan, Quart. Appl. Math. \textbf{32 }%
(1974)\textbf{\ }215.

\bibitem{9} S. Gasirowicz, "\textit{Quantum Mechanics", 3rd Edition (}John
Wiley and Sons, Inc., New Jersey 2003).
\end{thebibliography}
\end{document}